\documentclass[preprint,aps,showpacs,prl,amsmath]{revtex4-1}

\usepackage{amssymb}
\usepackage{graphicx}
\usepackage{dsfont}
\usepackage{color,soul}

\usepackage{subfig}
\usepackage{hyperref}
\usepackage{float}

\begin{document}

	\title{Optical Hall effect in strained graphene}
	
	\author{V. Hung Nguyen} \email{viet-hung.nguyen@uclouvain.be} \address{Institute of Condensed Matter and Nanosciences, Universit\'{e} catholique de Louvain, Chemin des \'{e}toiles 8, B-1348 Louvain-la-Neuve, Belgium}
	\author{A. Lherbier} \address{Institute of Condensed Matter and Nanosciences, Universit\'{e} catholique de Louvain, Chemin des \'{e}toiles 8, B-1348 Louvain-la-Neuve, Belgium}
	\author{J.-C. Charlier} \address{Institute of Condensed Matter and Nanosciences, Universit\'{e} catholique de Louvain, Chemin des \'{e}toiles 8, B-1348 Louvain-la-Neuve, Belgium}

\begin{abstract}

When passing an optical medium in the presence of a magnetic field, the polarization of light can be rotated either when reflected at the surface (Kerr effect) or when transmitted through the material (Faraday rotation). This phenomenon is a direct consequence of the optical Hall effect arising from the light-charge carrier interaction in solid state systems subjected to an external magnetic field, in analogy with the conventional Hall effect. The optical Hall effect has been explored in many thin films and also more recently in 2D layered materials. Here, an alternative approach based on strain engineering is proposed to achieve an optical Hall conductivity in graphene without magnetic field. Indeed, strain induces lattice symmetry breaking and hence can result in a finite optical Hall conductivity. First-principles calculations also predict this strain-induced optical Hall effect in other 2D materials. Combining with the possibility of tuning the light energy and polarization, the strain amplitude and direction, and the nature of the optical medium, large ranges of positive and negative optical Hall conductivities are predicted, thus opening the way to use these atomistic thin materials in novel specific opto-electro-mechanical devices.
\end{abstract}

\maketitle

The Hall effect \cite{hall79} is a fascinating phenomenon describing electrical conduction transverse to an applied electric field which is usually obtained thanks to a magnetic field. Its quantized version, the quantum Hall effect, has become one of the key tools for exploring quantum phenoma in 2D mesoscopic systems \cite{prgi90}. Although most of the works have concentrated on static properties, the optical Hall effect (OHE) is another exceptional feature \cite{schu16}. Indeed, while the longitudinal optical conductivity is related to the light absorption, a finite optical Hall component is the basis of the Faraday rotation and magneto-optic Kerr effect \cite{karg15}. Studying the OHE, on the one hand, is necessary to understand fully the picture of the dynamics of charges interacting with light and on the other hand, is a guide for magneto-optical applications, e.g., in optical diodes and other non-reciprocal optical elements \cite{lebi11}. Additionally, quantum Hall effect measurements in both DC and AC cases have been known as the basis of metrology applications \cite{jsch16}.  

In recent years, graphene and 2D layered materials have attracted increasing attention for many fundamental researches and applications \cite{ferr15}. Especially, due to its unusual electrical, optical properties and outstanding mechanical properties, graphene has been shown to be very promising for specific applications in flexible electronics \cite{jang16}, photonics and optoelectronics \cite{bona10}. In flexible electronics, the attractiveness of graphene lies in its excellent mechanical endurance and high sensitivity of the electronic properties to strain \cite{chzi16}. Due to its unconventional electronic structure with a linear dispersion at low energies, graphene has been widely used for numerous photonic and optoelectronic devices, operating in a broad spectral range from the ultraviolet, visible and near-infrared to the mid-infrared, far-infrared and even to the terahertz and microwave regions \cite{bona10}. Its applications include transparent electrodes, solar cells, optical modulators and photodetectors \cite{kopp14}.

Actually, the optical properties of graphene and related materials have been already reported in numerous published works \cite{grun03,nair08,stau08,yang09,wrig09,yang10,fmak11,yuan11,moon13,hale14,yang16}. The OHE in graphene subjected to an external magnetic field has been also theoretically and experimentally explored \cite{mori09,cras11,shim13,soun13,heym14,skul15,schi16,kuzm16}. The magnetic field breaks the time-reversal symmetry in graphene and hence, similarly to the static case, a finite optical Hall conductivity can be achieved. On this basis, the Faraday rotation of a few degrees in modest magnetic fields has been experimentally observed \cite{cras11,shim13}. It has been also shown that strain engineering is an efficient technique to modulate the optical properties of graphene \cite{pere10,pell10,dong14,gxni14,leyy15}. In particular, the strain can break the lattice symmetry and the optical spectrum of graphene exhibits strong anisotropy and dichroism. These findings appeal for an in-depth investigation of the optical Hall conductivity in strained graphene, which is still missing.

The aim of the present work is to investigate thoroughly the emergence of OHE in graphene systems subjected to strain, using the density functional theory (DFT) and parametrized tight-binding (TB) approaches \cite{supmat}. It is found that when strain is applied, a finite optical Hall conductivity is observed in graphene even though magnetic field is zero. Especially, this conductivity has rich properties, compared to the longitudinal component and conventional Hall effect obtained under an external magnetic field. In particular, the strain-induced Hall conductivity can be modulated while its sign can be reversed by tuning incident light (frequency and polarization) and/or strain (magnitude and direction). Finally, it is worth noting that this strain-induced OHE is demonstrated to be common for many other 2D materials and the explored properties could be the basis of several novel applications in opto-electro-mechanical systems. 

\section*{Results}

First-principles (at the DFT level) and parametrized TB approaches \cite{supmat} were employed to investigate the opto-electro-mechanical graphene systems schematized in Fig.1.a where a linearly polarized light with energy $\hbar \omega$ and polarization angle $\phi$ is considered and an in-plane uniaxial strain of magnitude $\epsilon$ is applied in the direction $\theta$ with respect to the armchair direction of graphene lattice.

In Figs.1.b-c, the optical conductivity components are computed as a function of light energy in monolayer graphene without and with strain. As seen, our parametrized TB calculations reproduce very well the DFT results at low and high energies. A slight discrepancy between two methods occurs only in the medium energy range where the conductivity peaks are present. In spite of this fact, the two methods are still in very good agreement for the investigation of the overall spectrum of optical conductivities in both unstrained and strained graphene systems (see the further demonstration in the Supplemental Material \cite{supmat}). Very remarkably,
the optical Hall conductivity is found to be zero for unstrained graphene but exhibits finite values when strain is applied. This is accompanied by a peak splitting in the longitudinal optical conductivity spectrum.
\begin{figure}[!t]
	\centering
	\includegraphics[width = 0.98\textwidth]{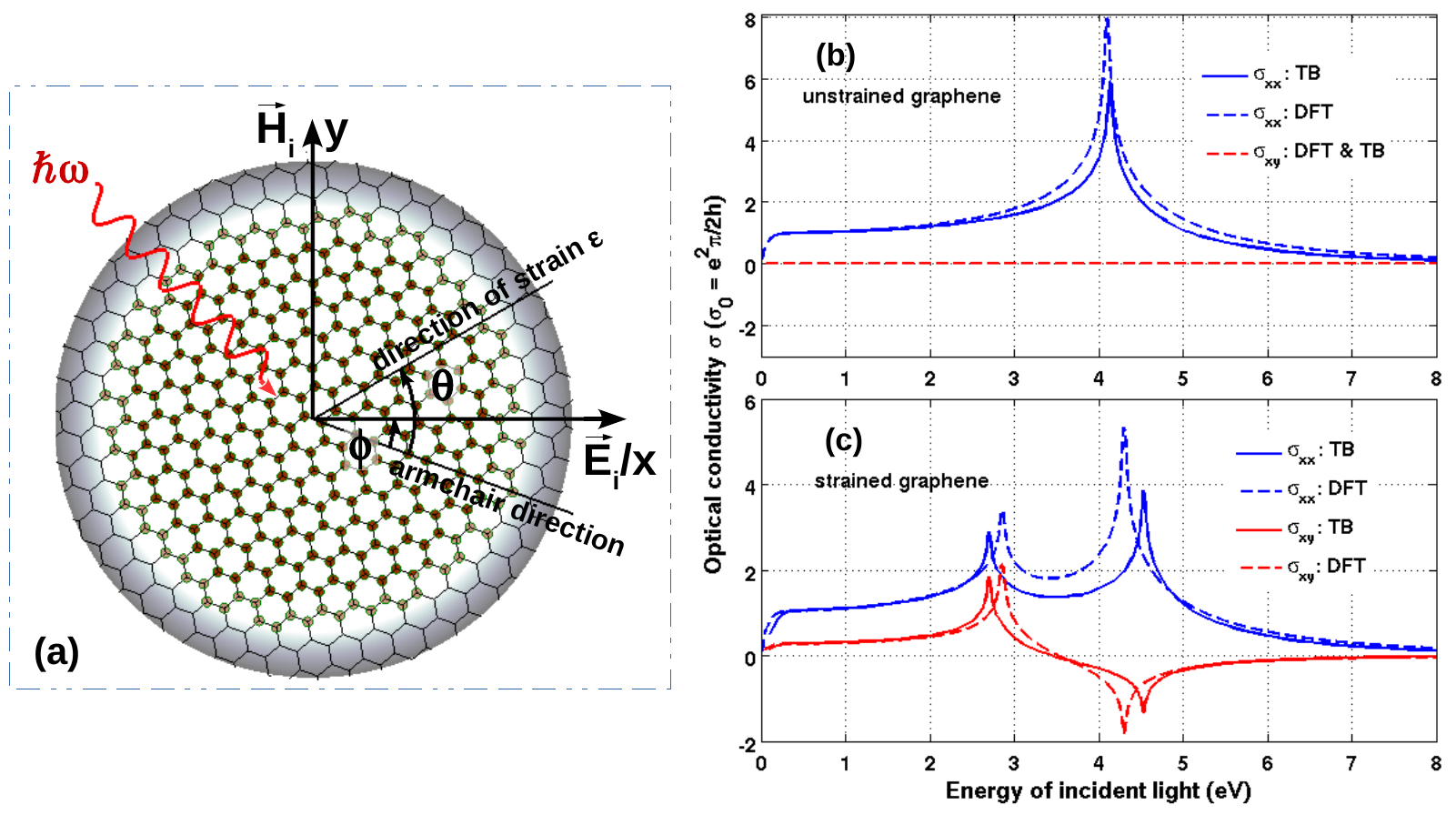} 
	\caption{Schematic representation of the opto-electro-mechanical graphene-based system (a). Optical conductivities, predicted by both DFT (dashed) and TB (solid lines) calculations, in graphene without (b) and with strain (c).}
	\label{fig_sim1}
\end{figure}

In order to understand the origins of the optical Hall effect observed, the effects of strain on the optical conductivities are investigated in more detail in Fig.2. Note that in the graphene bandstructure, there are six Dirac cones at the corners ($K$-points) of its Brillouin zone, however, they are characterized by only two distinguishable ones. Accordingly, six saddle ($M$-) points occur in the middle of these Dirac cones and only three are distinguishable. In the unstrained case, these saddle points are degenerate but can be separated in the energy when strain is applied (see the energy color map in Fig.2.a). Because the optical transitions of charges around the $\Gamma$-point is low (see in Figs.2.b-c and the related discussions below), the optical spectra of graphene can be basically understood by analyzing the transitions around $K$- and $M$- points. Indeed, the optical transitions around the $K$-points with a linear energy dispersion result in an almost flat spectrum of the longitudinal conductivity $\sigma_{xx}$ with a universal value of $\sigma_0 = e^2/4\hbar$ at low energies for monolayer graphene \cite{fmak11,yuan11,moon13,hale14} (see Fig.2.d). At moderately high energies when the transitions around the $M$-points take place, this conductivity exhibits a relatively high peak, which is essentially due to high optical transitions and high density of states of graphene at these points. When strain is applied, two main features occur. First, strain effects make graphene lattice very anisotropic, leading to a strong anisotropy of the absorption spectrum, i.e., the conductivity $\sigma_{xx}$ is strongly dependent on the light polarization and strain direction (see the further details in \cite{pere10,pell10}). Second, as already mentioned, the strain can break the degeneracy of saddle points, leading to the separation of the peaks of $\sigma_{xx}$ as presented in Fig.1.c and Figs.2.d.
\begin{figure}[!b]
	\centering
	\includegraphics[width = 1.0\textwidth]{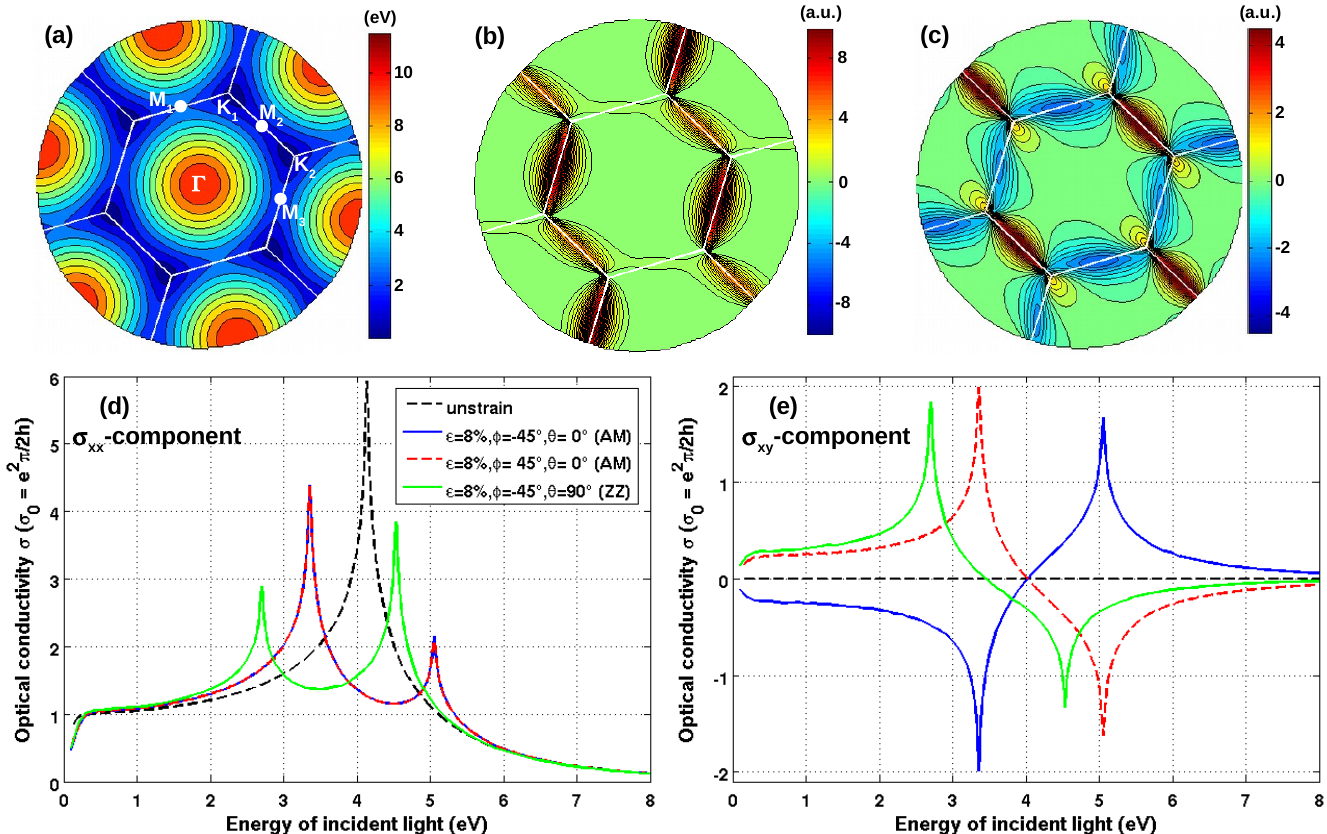} 
	\caption{Conduction bands (a), transition coefficients $P_{xx}$($\vec k)$ (b) and $P_{xy}$($\vec k)$ (c) (see text) are presented for light polarization $\phi = -45^\circ$ and uniaxial strain $(\epsilon,\theta) = (8\%,75^\circ)$. Longitudinal optical conductivity (d) and optical Hall component (e) in uniaxial strained graphene obtained for different $\phi$ and $\theta$ angles.}
	\label{fig_sim2}
\end{figure}

Now, the OHE in strained graphene is explored, i.e., a finite strain-induced conductivity $\sigma_{xy}$ as displayed in Fig.1.c and Fig.2.e. One of the key terms to determine the conductivity $\sigma_{pq}$ from the Kubo formula (see the details in \cite{supmat}), and hence understand the underlying mechanism if this OHE, is the optical transition functions $P_{pq}(\vec k) = \left\langle s_v \right| \hat{v}_p \left| s_c \right\rangle \left\langle s_c \right| \hat{v}_q \left| s_v \right\rangle$. Here, $\left| s_{v,c} \right\rangle$ are eigenstates in the valence/conduction bands, respectively, and $\hat{v}_{pq}$ ($p,q = x,y$) are the velocity operators. Different from $P_{xx} (\vec k)$ that is a non-negative function (see Fig.2.b), $P_{xy}(\vec k)$ is either positive or negative when considering the whole Brillouin zone of graphene (see in Fig.2.c). The Hall conductivity can be hence rewritten as $\sigma_{xy} = \sigma_{xy}^+ - \sigma_{xy}^-$ where $\sigma_{xy}^\pm$ are the absolute values of terms corresponding to positive and negative $P_{xy}(\vec k)$, respectively. In unstrained case, $\sigma_{xy}^+ = \sigma_{xy}^-$ and hence the conductivity $\sigma_{xy}$ is totally zero for any polarization and energy of light. In strained graphene, the equality of $\sigma_{xy}^\pm$ is broken by strain, leading to a finite $\sigma_{xy}$. Physically, it can be understood that in analogy to the magnetic-field effects that break the time-reversal symmetry, the lattice symmetry breaking by strain changes the optical responses and hence results in the OHE in graphene.

For all cases presented in Fig.2.e, two interesting properties of $\sigma_{xy}$ are found. First, similarly to the conductivity $\sigma_{xx}$, the Hall conductivity $\sigma_{xy}$ exhibits an almost flat spectrum in the low energy regime and tends to zero at very high energies. Second, high peaks of $\sigma_{xy}$ are also observed, especially, at the same energies as for the $\sigma_{xx}$-component (see Fig.1.c and Figs.2.d-e). However, while $\sigma_{xx}$ is always positive, the full spectrum of $\sigma_{xy}$ presents both positive and negative peaks and accordingly, two (low and high) energy regimes where this Hall conductivity has opposite signs. This novel property can be understood as follows. As mentioned, $P_{xy}(\vec k)$ can be either positive or negative in specific areas of the Brillouin zone of graphene. Simultaneously, when strain is applied, the graphene bandstructure is deformed, leading to the separation of degenerate bands in such different areas. Because of these two features, two terms $\sigma_{xy}^\pm$ are alternatively dominant in different energy regimes, leading to opposite Hall conductivities at low and high energies as observed. Additionally, this Hall conductivity is also predicted to be strongly anisotropic (see the detailed discussions below), similar to the $\sigma_{xx}$ component \cite{pere10,pell10}.
\begin{figure}[!t]
	\centering
	\includegraphics[width = 0.65\textwidth]{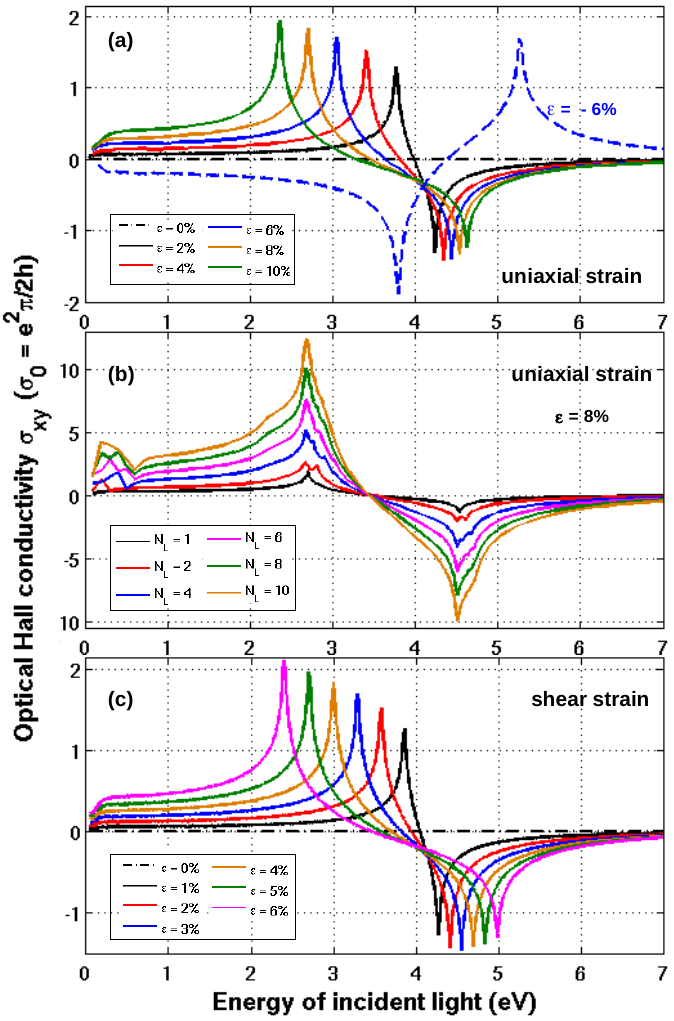} 
	\caption{Optical Hall conductivity in monolayer graphene with different strains (a,c) and in few-layer systems (b). Uniaxial ($\theta = 90^\circ$)/shear ($\theta = 45^\circ$) strains are applied in (a,b)/(c), respectively. The light polarization is fixed to $\phi = -45^\circ$.}
	\label{fig_sim3}
\end{figure}

In the following, the possibilities of achieving and tuning a large Hall conductivity are investigated. First, the Hall conductivity obtained for different uniaxial strains is displayed in Fig.3.a. The light polarization $\phi = -45^\circ$ is chosen so as to achieve the largest $\sigma_{xy}$ for strains applied along the direction $\theta = 90^\circ$ (see Fig.4 below). At low energies, the Hall conductivity is gradually increased when increasing the strain magnitude and, particularly, reaches $\sim 0.5 \sigma_0$ for $\epsilon = 10\%$ and $\hbar \omega = 1.5$ eV. At high energies, $\sigma_{xy}$-peaks occur and are very rapidly increased for low strains, i.e., they reach values larger than $\sigma_0$ for a small strain of only $2\%$, and progressively saturate at large strains. 
Basically, the high peaks of $\sigma_{xy}$ $\sim 1 \div 2$ $\sigma_0$ can be achieved for a strain of only a few percents. In order to further enlarge $\sigma_{xy}$ (e.g., to achieve large Faraday/Kerr rotations), another strategy is suggested \cite{fmak11}, i.e., to use few-layer graphene systems. Indeed, as presented in Fig.3.b, the Hall conductivity is almost linearly increased as a function of number of graphene layers $N_L$. In particular, extremely large peaks of $\sigma_{xy}$, $\sim$13$\sigma_0$ at $\hbar \omega \simeq 2.67$ eV and $\sim$-10$\sigma_0$ at $4.51$ eV, and a large value $\sim$5$\sigma_0$ at 2 eV are achieved for a uniaxial strain of 8$\%$ and $N_L = 10$. 
\begin{figure*}[!t]
	\centering
	\includegraphics[width = 0.95\textwidth]{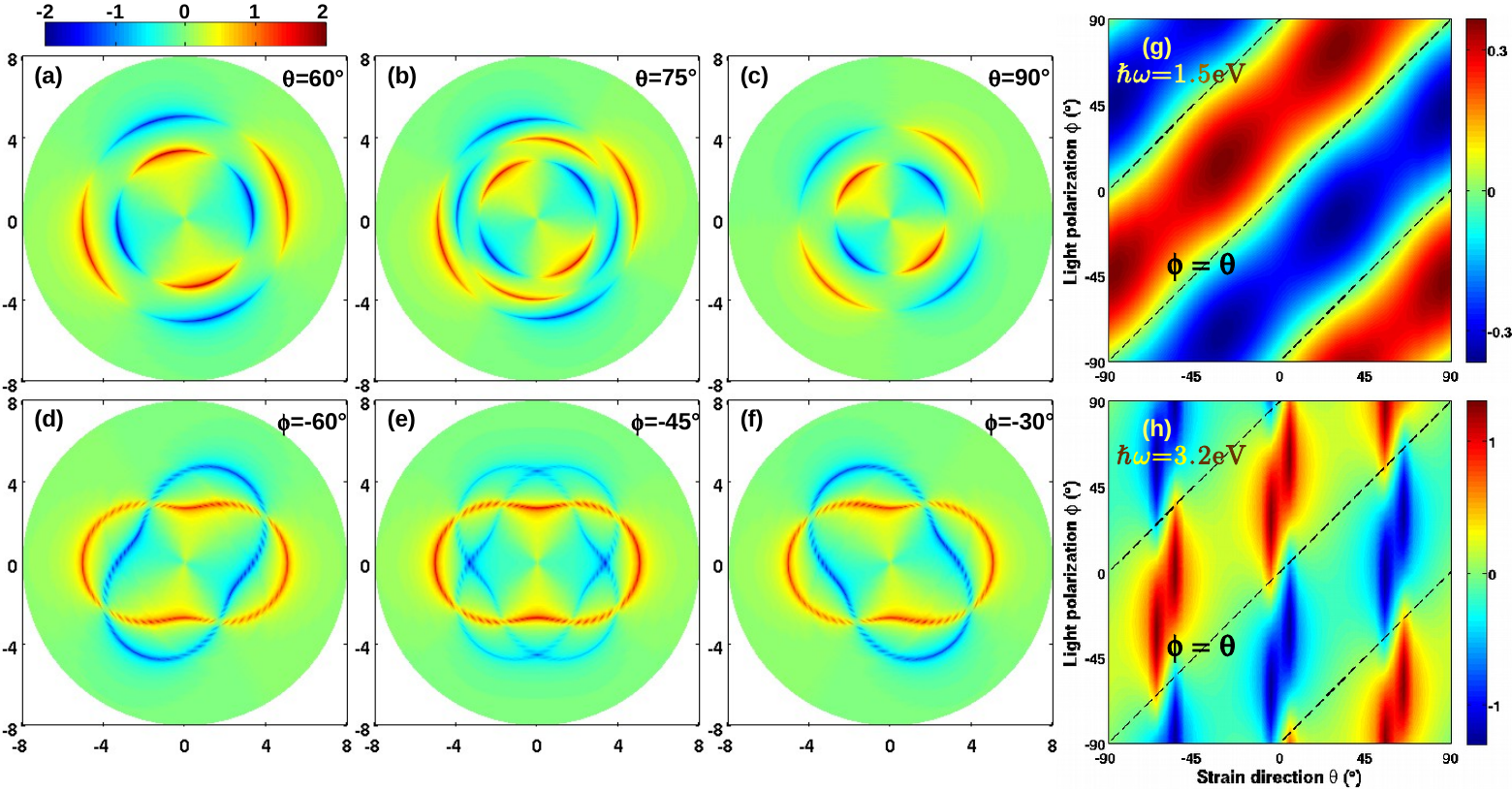} 
	\caption{Optical Hall conductivity in monolayer graphene with respect to the light energy $\hbar\omega$ ($> 0$ and in eV) and polarization $\phi$, and the strain direction $\theta$. The polar axis of (a-d) presents the energy $\hbar\omega$ while the azimuthal direction indicates the angles $\phi$ in (a-b) and $\theta$ in (c-d). The dashed lines in (e,f) indicate the cases $\phi - \theta = n\pi/2$. The amplitude of uniaxial strain is fixed to $\epsilon = 8\%$.}
	\label{fig_sim4}
\end{figure*}

The effect of shear strains is also investigated in Fig.3.c. Interestingly, a smaller shear strain than uniaxial one is required in order to achieve a similarly large $\sigma_{xy}$. Our calculations show that overall, the amplitude of Hall conductivity is proportional to the off-diagonal term $\epsilon_{xy}$ of the strain tensor \cite{supmat}. In the Oxy axis chosen as in Fig.1.a, $\epsilon_{xy} = (1+\alpha) \textrm{sin}(2(\theta-\phi)) \epsilon/2$ and $\textrm{cos}(2(\theta-\phi)) \epsilon$ for uniaxial and shear strains, respectively, with the Poisson ratio $\alpha = 0.165$ \cite{blak70}. On this basis, the largest $\sigma_{xy}$ is respectively proportional to $(1+\alpha)\epsilon_{uniaxial}/2$ and $\epsilon_{shear}$, i.e., with two strains satisfying $\epsilon_{shear}/\epsilon_{uniaxial} \simeq (1+\alpha)/2 = 0.5825$ the similar $\sigma_{xy}$-magnitude can be achieved. 

Achieving both positive and negative values of the Hall conductivity is a very novel/ promising result for practical applications (see the discussions below). As presented in Fig.2 and Fig.3, the sign of $\sigma_{xy}$ can be reversed by tuning the light energy. In Fig.3.a, another possibility of reversing this conductivity is also found, i.e., by switching from positive to negative strain.

As already shown in Fig.2, the Hall conductivity induced by strain is predicted to be strongly anisotropic, i.e., strongly
dependent on the light polarization and strain direction. In Fig.4, colormaps showing the full dependence of $\sigma_{xy}$ on the directions of light polarization and strain are presented. For a given strain (see Figs.4.a-c and 4.g-h), the $\phi$-dependence of $\sigma_{xy}$ generally satisfies the simple rule $\sigma_{xy} \varpropto \textrm{sin}(2(\phi - \theta - \theta_s))$ where $\theta_s$ is a function of $\omega$, $\epsilon$ and $\theta$. At low energies, $\theta_s$ is approximately zero for all strain directions. For a given light polarization (see Figs.4.d-h), the $\theta$-dependence of $\sigma_{xy}$ also satisfies the same rule at low energies but due to the presence of $\sigma_{xy}$-peaks, becomes more complex at high energies. Additionally, the peaks of $\sigma_{xy}$ in ($\omega,\theta$)-maps presents three specific peanut lines with the symmetry under a rotation of 60$^\circ$ (see in Figs.4.d-f). The full spectrum of $\sigma_{xy}$ generally presents three peaks but only two peaks are observed if the light polarization or strain direction is parallel to the armchair or zigzag directions of graphene (see Fig.2, Figs.4.a, 4.c, 4.d, and 4.f). Actually, when uniaxial strain is applied, there are two symmetry directions of lattice deformation, i.e., either parallel or perpendicular to the strain direction. The most common property observed here is that $\sigma_{xy}$ is generally high if $\phi - \theta \simeq n\pi/2 + \pi/4$ and low if the light polarization is parallel to any symmetry direction of lattice deformation mentioned above, i.e., $\phi - \theta \simeq n\pi/2$ as seen in Figs.4.g-h. Therefore, the overall dependence of $\sigma_{xy}$ on $\phi$ and $\theta$ generally satisfies the rule $\sigma_{xy} \varpropto \textrm{sin}(2(\phi - \theta))$ although it is considerably disturbed by the presence of $\sigma_{xy}$-peaks at high energies (see Fig.4.h). For shear strains, this rule becomes $\sigma_{xy} \varpropto \textrm{cos}(2(\phi - \theta))$ as discussed above for Fig.3.c. Most interestingly, the direction dependence explored here suggests other possibilities of controlling the sign of $\sigma_{xy}$, i.e., by changing the light polarization and strain direction.

\begin{figure*}[b!]
	\centering
	\includegraphics[width = 0.85\textwidth]{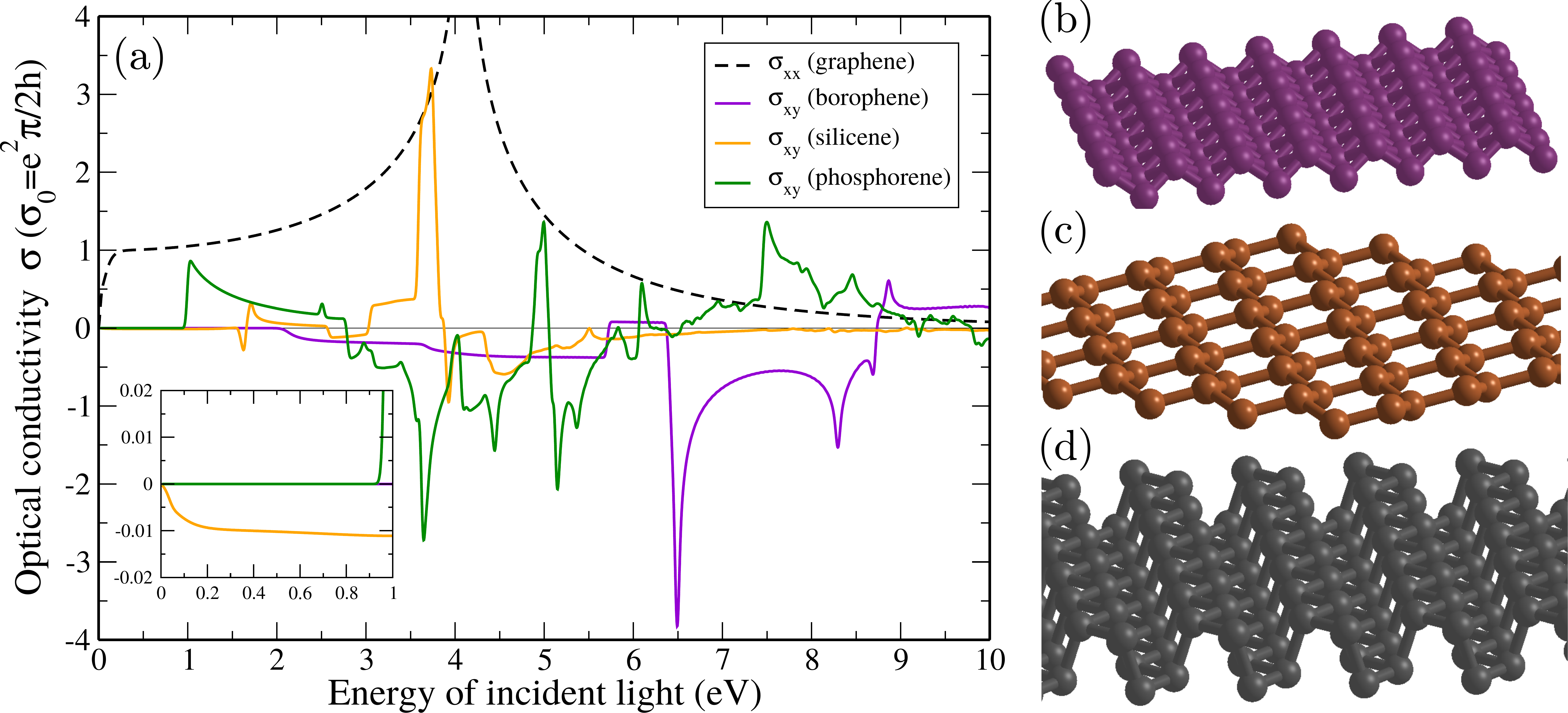} 
	\caption{Strain-induced optical Hall effect (a) in borophene (b), silicene (c), and phosphorene (d). A uniaxial strain of $4\%$ is applied in the direction $\theta = 0^\circ$ while the light polarization is $\phi = 45^\circ$. The inset in (a) is a zoom of the optical Hall conductivities at low energies.}
	\label{fig_sim5}
\end{figure*}
Finally, it is worth noting that in analogy to the effect of an external magnetic field, the present opto-electro-mechanical effect is essentially due to the lattice symmetry breaking and hence completely general, i.e., it can be observed in many other 2D materials even though they are metallic, semimetallic or semiconducting. As some examples, ab initio optical Hall conductivities calculated in silicene, borophene, and phosphorene are presented in Fig.5, which demonstrates that the mechanism explored in this study is indeed very general. Note however, that the optical Hall conductivity remains zero until the optical gap is reached. For instance, while borophene is a metal, optical transitions occur only for energies above 2--3 eV (depending on the applied strain)\cite{lher16}. For phosphorene, the energy threshold for transitions is $\sim$1 eV at the level of DFT but the optical gap can be modified because of screening and exitonic effects. Finally, due to its similar bandstructure at low energies, there is no gap for the optical Hall conductivity in silicene (see the inset of Fig.5), but its value is much smaller than that obtained in graphene. The latter might be explained by a lower Fermi velocity in silicene compared to graphene as the optical transitions are proportional to the expectation value of velocity operators \cite{supmat}.

\section*{Discussion}

Now, let us discuss novel properties of this strain-induced OHE and related possible applications. First, a finite $\sigma_{xy}$ can be achieved under zero magnetic field. Second, different from the longitudinal component, the sign of this conductivity can be reversed by tuning the light energy and polarization, changing the strain type (i.e., from positive to negative) and its direction. Note that the dependence on light polarization can not be observed for $\sigma_{xy}$ induced by an external magnetic field where it is totally isotropic \cite{mori09,cras11,shim13}. These switching possibilities can be exploited to explore novel applications in opto-electro-mechanical devices. Third, extremely large $\sigma_{xy}$ can be observed when shining on the system by a light beam with appropriate energy and can be further enlarged using few-layer systems, of course, at the expense of higher transmittance loss. This is an important ingredient for designing efficient Faraday rotators and related applications. 

\begin{figure}[b!]
	\captionsetup[subfigure]{labelformat=empty}
	\begin{minipage}{0.95\textwidth}
		\centering
		\subfloat[]{\includegraphics[width=0.65\columnwidth]{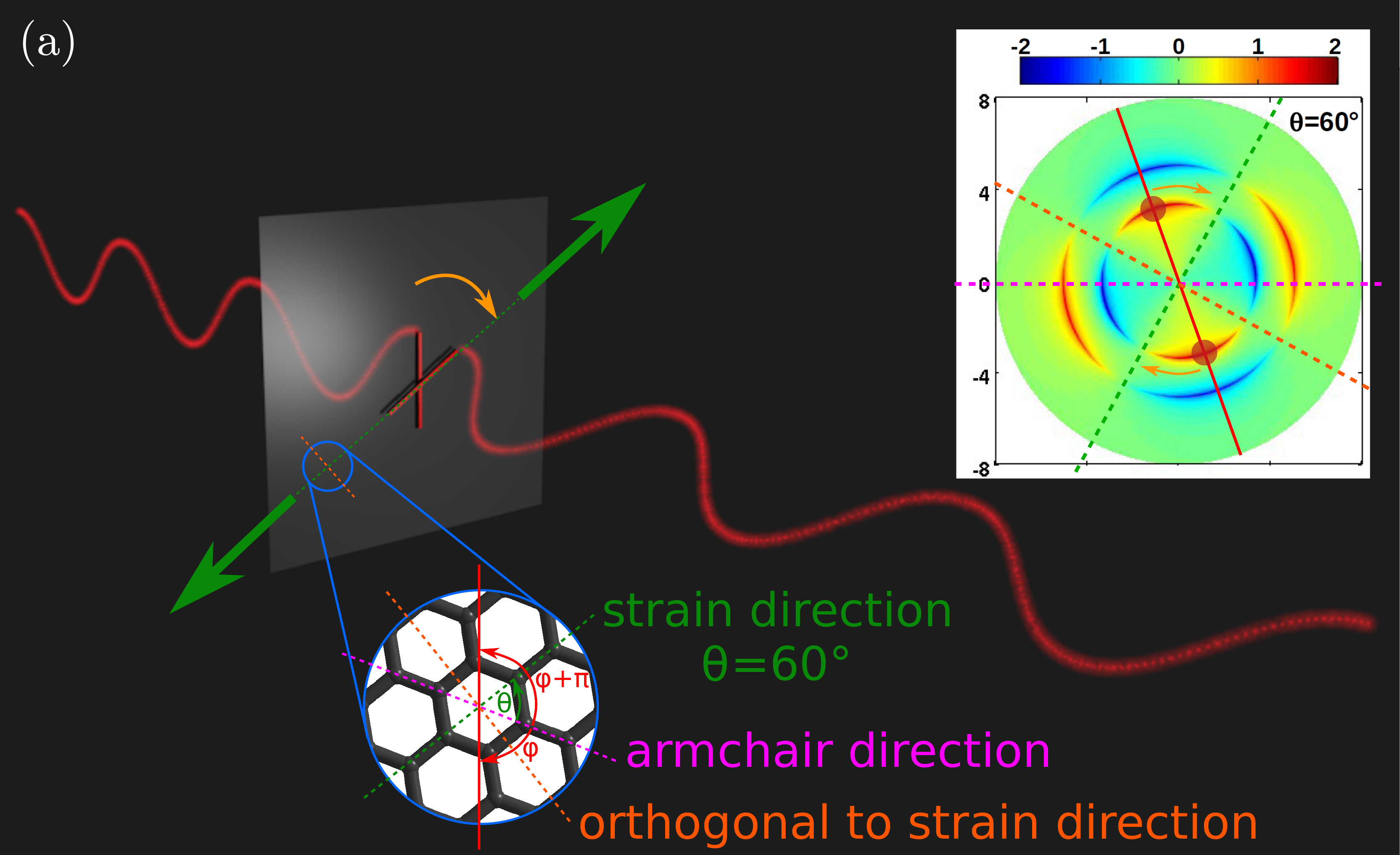}}
	\end{minipage}\par
	\begin{minipage}{0.95\textwidth}
		\centering
		\subfloat[]{\includegraphics[width=0.65\columnwidth]{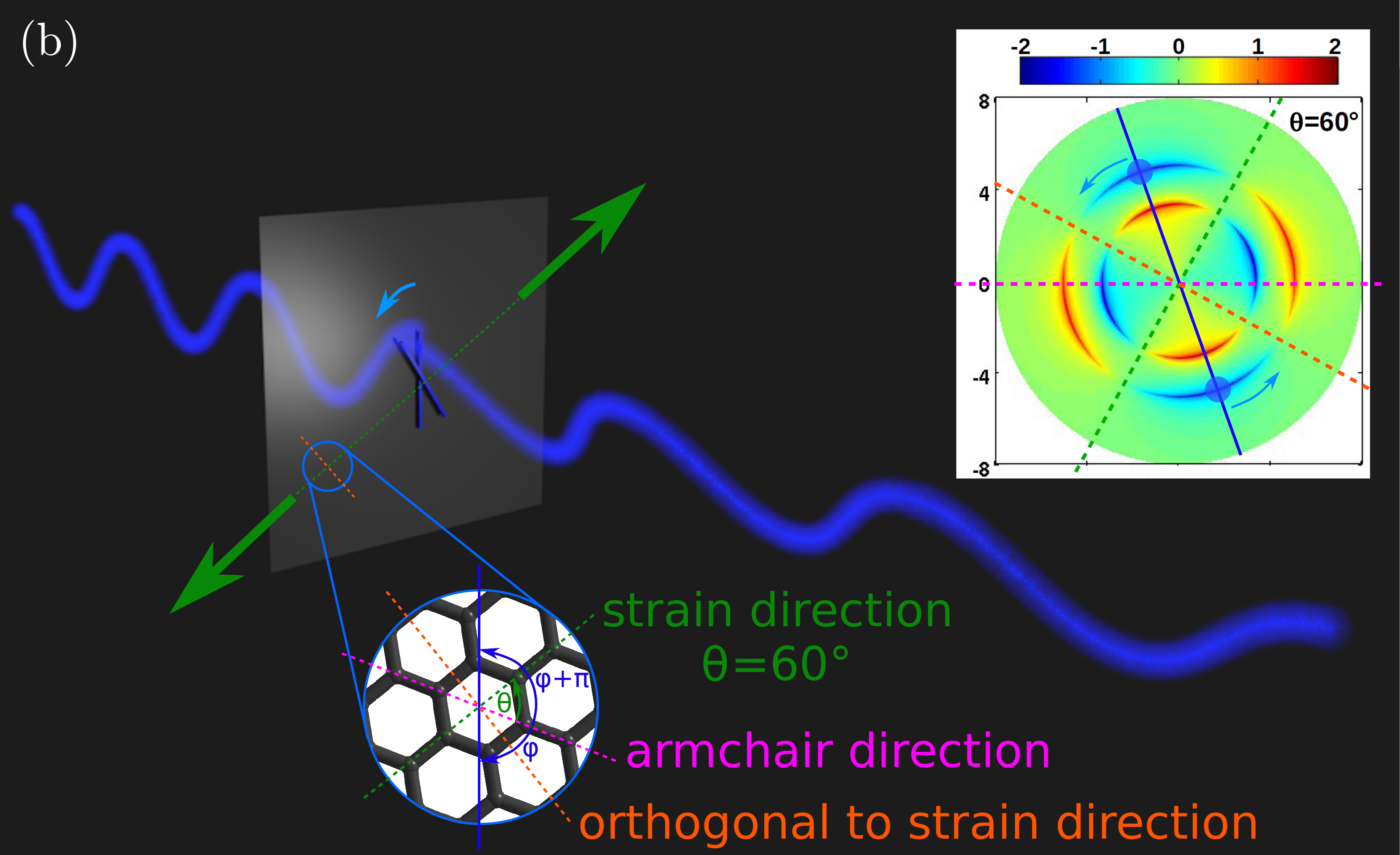}}
	\end{minipage}
	\caption{Faraday rotation with low (a) and high (b) energies of the incident light.}
	\label{supfig2}
\end{figure}
Interestingly, with the binary property of $\sigma_{xy}$ when varying the ligh energy and its polarization, the designed Faraday rotator can work as an optical polarizer with a specific output polarization that is dependent on and can be controlled by tuning the strain direction. Indeed, Fig.6 illustrates this feature in the case of strain angle $\theta=60^\circ$. The linearly polarized light carries an electric field oscillating along a given axis determined by the angle $\phi$ (see in Fig.1 and the red line in Fig.6.a). In the low energy regime and for the chosen angle $\phi$ as in the Fig.6.a, the corresponding value of the Hall conductivity is positive (see inset) and therefore the light rotates in the clockwise direction (orange arrow) towards the axis determined by the strain direction (green axis). When its polarization is aligned with the strain direction, $\sigma_{xy}$ is zero and hence the light does not rotate anymore. It is worth noting that due to the dependence of $\sigma_{xy}$ (i.e., both intensity and sign) on the light polarization $\phi$ as presented in Figs.4.a-c and more clearly in Fig.4.g, all the light beams in the low energy regime are always rotated towards the strain axis either with a clockwise or anti-clockwise rotation. Note that by symmetry, the two directions $\phi$ and $\phi + 180^\circ$ are equivalent. The direction orthogonal to the strain (dark orange axis) presents also $\sigma_{xy}=0$ but corresponds to a saddle point since for any small deviation, the light rotates away from this point. Such a device can thus be used as an efficient light polarizer in the sense that it does not only filter polarized light but also convert light that is not correctly oriented into the desired orientation determined by the direction of applied strain.

Additionally, if one now considers linearly polarized light with several frequencies, it is possible to separate the spectrum into two parts. As discussed above, the low energy components of the incoming light rotate towards the strain axis. In the high energy regime the optical Hall conductivity has a opposite sign, compared to that obtained at low energies. Therefore, as illustrated in Fig.6.b, the high energy components rotate towards the axis orthogonal to the strain direction. Thus, the incoming light can be separated into two beams having orthogonal polarizations with the energy threshold ($\sim$ 3-4 eV depending on the applied strain), which is the energy point where the sign of $\sigma_{xy}(\omega)$ is reversed.

With these properties discussed above, the system can also act as a converter of circularly polarized light into a linearly polarized light. Actually, a circularly polarized light can be considered as a combination of two orthogonal and dephased linearly polarized lights. Since these two linearly polarized lights can be subjected to opposite Faraday rotations, the transmitted light becomes linearly polarized along to the strain (resp. its orthogonal) axis in the case of low (resp. high) energies, respectively, with an amplified electric field (see the detailed illustration in \cite{supmat}).

Finally, it is worth noting that since the transmittance is extremely high in the considered thin 2D systems \cite{lher16}, the intensity of transmitted light maintains high values compared to the incoming one, which is different from 3D systems where a large part of light can be reflected or absorbed. 

To conclude, strain engineering has been demonstrated to be a novel and alternative approach to efficiently generate optical Hall effect in graphene and 2D materials. Compared to the conventional effect observed under an external magnetic field, the strain-induced optical Hall conductivity exhibits rich properties, i.e., its value can be modulated whereas its sign can be reversed by tuning incident light (frequency and polarization) and/or strain (magnitude and direction). The observed properties could be exploited to explore novel optical devices and, particularly, specific applications in opto-electro-mechanical 2D systems.

\section*{Methods}

\textbf{First-principles calculations}. First-principles calculations were performed using the self-consistent density functional theory (DFT) within the GGA-PBE approach implemented in the SIESTA \cite{sole02} package. The complex optical conductivity tensor $\sigma_{pq}$ is derived from the complex dielectric tensor $\varepsilon_{pq}$ using the formula $\sigma_{pq}(\omega)=-i\omega \varepsilon_0 \varepsilon_{pq}(\omega)$, where $\varepsilon_0$ is the vacuum dielectric constant. Note that the calculation of $\varepsilon_{pq}$ requires a specific treatment because of the 2D nature of the system. Further details can be found in the supplementary materials \cite{supmat}.

\textbf{Tight-binding calculations}. The tight binding (TB) calculations for graphene were performed within a third-nearest-neighbors orthogonal model that has been demonstrated to give reasonably accurate results, compared to the DFT ones \cite{lher12}. In this work, the parameters (i.e., onsite energy and three hopping terms) of this TB model were elaborated from the DFT data so as to achieve the best agreement between the two methods in the calculation of the electronic bandstructure and the optical conductivities. The TB optical conductivities are computed using the standard Kubo formula \cite{moon13,hale14}. When strain is applied, the graphene lattice is deformed and the $C-C$ bond length is hence modified. Taking into account this effect, the TB hopping energies are modified following the exponential decay law as in \cite{pere09}. The detailed description of this calculation method can be also found in the supplementary materials \cite{supmat}.

\section*{Acknowledgements}
The authors would like to thank Prof. Beno\^{i}t Hackens (UCL) for fruitful comments and discussions. The work was supported by the F.R.S.-FNRS of Belgium through the research project (N$^\circ$ T.1077.15), by the F\'{e}d\'{e}ration Wallonie-Bruxelles through the ARC entitled \textit{3D Nanoarchitecturing of 2D crystals} (N$^\circ$ 16/21-077) and by the European Union's Horizon 2020 research and innovation programme (N$^\circ$ 696656). Computational resources have been provided by the supercomputing facilities of the Universit\'{e} catholique de Louvain (CISM/UCL) and the Consortium des Equipements de Calcul Intensif en F\'{e}d\'{e}ration Wallonie Bruxelles (CECI) funded by the F. R. S.-FNRS under the convention N$^\circ$ 2.5020.11.
\section*{Author contributions}
V.H.N and A.L. performed the calculations. All authors discussed the results and wrote the manuscript.
\section*{Additional information}
\textbf{Competing financial interests}: The authors declare no competing financial interests.

\newpage

\begin{center}
\textbf{Supplementary Material of "Optical Hall effect in strained graphene"}

V. Hung Nguyen, A. Lherbier, and J.-C. Charlier

\textit{Institute of Condensed Matter and Nanosciences, Universit\'{e} catholique de Louvain, Chemin des \'{e}toiles 8, B-1348 Louvain-la-Neuve, Belgium}
\end{center}

\pagenumbering{arabic}

\section{I. Optical Hall conductivity}

The optical Hall conductivities presented in the main text were computed using either a first-principles ab initio method or a parameterized tight-binding approach.

\subsection{1. First-principles calculations}

First-principles calculations were performed using the self-consistent density functional theory (DFT) method implemented in the SIESTA $[1]$ package. The exchange-correlation energy and electron-ion interaction are described using GGA-PBE $[2]$ functional and norm-conserving pseudopotentials $[3]$ in the fully non local form, respectively. A double-$\zeta$ polarized basis set of numerical atomic orbitals is used and the energy cutoff for real-space mesh is set to 500 $R_y$. The energy levels was populated following a Fermi-Dirac distribution function with an electronic temperature of 300 K. An inter-layer (vacuum) distance of 30 \AA~ was applied in order to avoid next cell image interactions. The calculations were performed for strained systems that, particularly for graphene, are obtained by applying the strain tensor described below on the pristine (i.e., relaxed) lattice. These calculations were performed in the primitive unit cell with a Monkhorst-Pack k point grid of 300$\times$300$\times$1 and for the optical calculation a mesh of 1100$\times$1100$\times$1 with a broadening factor of 25 meV.

The complex optical conductivity tensor $\sigma_{pq}(\omega)$ is derived from the complex dielectric tensor $\varepsilon_{pq}(\omega)$ using the formula $\sigma_{pq}(\omega)=-i\omega \varepsilon_0 \varepsilon_{pq}(\omega)$, where $\varepsilon_0$ is the vacuum dielectric constant. The real part of the optical conductivity is therefore related to the imaginary part of the dielectric tensor.
As in SIESTA package (version 3.1) the off-diagonal tensor elements ($p\neq q$) are not coded, we modified the code to also compute these terms and thus to obtain the optical Hall conductivities. One also notes that the dielectric tensor is computed for the case of a triply periodic simulation cell. Therefore, one has to renormalize the obtained 3D conductivities into 2D conductivities by using
a multiplying factor corresponding to the height of the simulation box (in our case 30 \AA~). That corresponds to the transformation of a 3D electron density $n^{\rm{3D}}=\frac{\# e^{-}}{V}$, where $V$ is the volume, into a 2D electron density $n^{\rm{2D}} = n^{\rm{3D}}\times H = \frac{\# e^{-}}{S}$, where $H$ is the simulation box height (vacuum distance between graphene layers),
and $S$ is the surface area of the simulation box corresponding to graphene.

\subsection{2. Tight-binding calculations}

\textbf{\textit{Tight binding models.}} To compute the electronic properties of graphene, most tight binding (TB) studies use a first-nearest-neighbors $\pi-\pi$* model, however, this model fails to describe the energy bands at high energies. In particular, it produces a totally symmetric band structure that is a significant discrepancy in comparison with the DFT one. In this regard, it has been shown $[4]$ that a third-nearest-neighbors TB model gives much more reasonable results since it allows to recover the existing asymmetry between valence ($\pi$) and conduction ($\pi$*) bands. The parameters of this model are basically composed of a single onsite term $\mu_{0}$ and three hopping terms $\nu_0^1$, $\nu_0^2$, and $\nu_0^2$ corresponding, respectively, to first, second, and third nearest neighbors. The monolayer graphene Hamiltonian hence reads as
\begin{eqnarray}
H_{1L} = \sum\limits_{i}^{} \mu_0 c_i^\dag c_i  + \sum\limits_{i, \left\langle j,k,l \right\rangle }^{} \left( \nu_0^1 c_i^\dag c_j + \nu_0^2 c_i^\dag c_k + \nu_0^3 c_i^\dag c_l \right)
\end{eqnarray}
In order to achieve the best consistency between the DFT and TB bandstructures in a large energy range (see Fig.S1.a), these TB parameters have been slightly modified, compared to those presented in $[4]$. In particular, we employed $\mu_0 = 654.4$ meV and $\nu_0^1 = -2845.8$ meV, $\nu_0^2 = 219.1$ meV, $\nu_0^3 = -259.4$ meV. 

Additionally, to investigate few-layer graphene systems, the interlayer coupling term
\begin{eqnarray}
H_{int} = \sum\limits_{i,j}^{} \left( \gamma_{ij} c_{i,1}^\dag c_{j,2} + h.c. \right)
\end{eqnarray}
was added. To obtain the best agreement with the DFT calculations (see Fig.S1.b), only interlayer couplings as in $[5]$ were taken into account, in particular, $\gamma_1 = 253$ meV and $\gamma_3 = \gamma_4 = 195.5$ meV.

\begin{center}
	\begin{minipage}{1.0\textwidth}
		\centering
		\includegraphics[width=0.62\columnwidth]{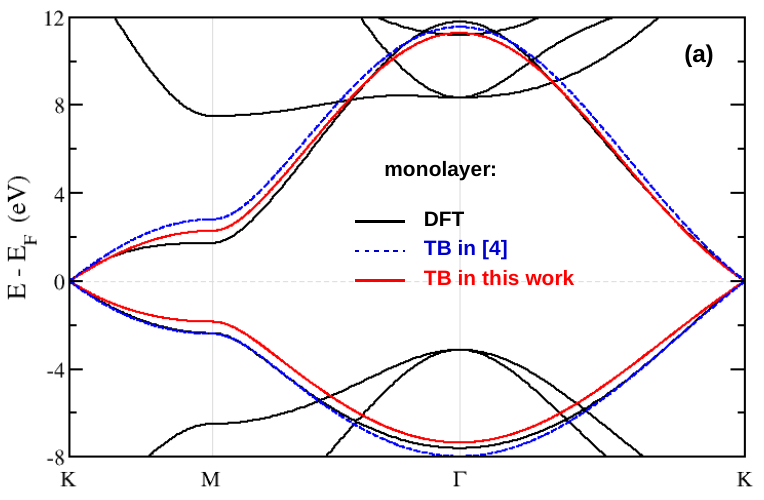}
		\includegraphics[width=0.62\columnwidth]{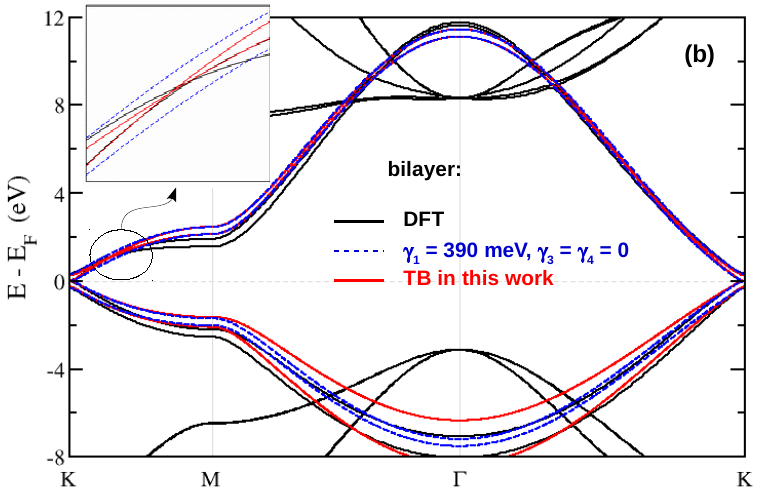}
	\end{minipage}\par
	
	\textbf{Fig. S1}: Comparison between DFT and parameterized TB calculations of graphene bandstructure: graphene monolayer (a) and bilayer (b).
\end{center}

\textbf{\textit{Strain effects in graphene}}. We investigate the effects of in-plane strain applied in the direction of the angle $\theta$ with respect to the armchair direction of graphene (see Fig.1 in the main text). This strain causes changes in the $C-C$ bond vector $\vec{r}_{ij}$ according to
\begin{equation}
\vec{r} =  \left\{ {\mathds{1} + {M_s}\left( \epsilon \right)} \right\}	\left[ {\begin{array}{*{20}{c}}
	{r_{||}}\\ {r_{\bot}}
	\end{array}} \right]_{\epsilon = 0} 
\end{equation}
where, $r_{||}$ ($r_{\bot}$) is the parallel (perpendicular) component of $\vec r$ with respect to the strain direction whereas the strain tensor is determined $[6]$ by
\begin{eqnarray}
M_{s} &=& \epsilon \left[ {\begin{array}{*{20}{c}}
	{1}&{0}\\ {0}&{-\alpha}
	\end{array}} \right]_{\{||,\bot\}} \leftrightarrow \epsilon \left[ {\begin{array}{*{20}{c}}
	{\textrm{cos}^2(\theta-\phi)-\alpha\textrm{sin}^2(\theta-\phi)}&{(1+\alpha)\textrm{cos}(\theta-\phi)\textrm{sin}(\theta-\phi)}\\ {(1+\alpha)\textrm{cos}(\theta-\phi)\textrm{sin}(\theta-\phi)}&{\textrm{sin}^2(\theta-\phi)-\alpha\textrm{cos}^2(\theta-\phi)}
	\end{array}} \right]_{\{x,y\}} \nonumber
\end{eqnarray}
for uniaxial strains and
\begin{eqnarray}
\hspace*{-4.3cm}
M_{s} &=& \epsilon \left[ {\begin{array}{*{20}{c}}
	{0}&{1}\\ {1}&{0}
	\end{array}} \right]_{\{||,\bot\}} \,\,\,\,\,\, \leftrightarrow \epsilon \left[ {\begin{array}{*{20}{c}}
	{\textrm{sin}(2(\phi - \theta))}&{\textrm{cos}(2(\theta-\phi))}\\ {\textrm{cos}(2(\theta-\phi))}&{\textrm{sin}(2(\theta - \phi))}
	\end{array}} \right]_{\{x,y\}} \nonumber
\end{eqnarray}
for shear strains, with the strain magnitude $\epsilon$ and Poisson ratio $\alpha = 0.165$ $[7]$. Here, $\{...,...\}$ represents the basis axis, the angles $\theta$ and $\phi$ are determined in Fig.1.a of the main text. Accordingly, the TB hopping energies under deformation are computed as
\begin{equation}
\nu_{ij} = \nu_{ij}^0 \exp\left\lbrace \beta \left( r_{ij}(0) - r_{ij}(\epsilon)\right) \right\rbrace 
\end{equation}
In order to achieve the best agreement with the DFT calculations, $\beta \simeq 16.9$ $nm^{-1}$ was used.

\textbf{\textit{Optical conductivities}}. The TB models described above were employed to compute the optical conductivities using the standard Kubo formula $[8,9]$:
\begin{eqnarray}
\sigma_{pq} (\omega) = \frac{2e^2\hbar}{iS}\sum\limits_{\textrm{\textbf{k}} \in \textrm{\textbf{BZ}}}^{}\sum\limits_{n,m}^{}\frac{f(E_n) - f(E_m)}{E_n(\textrm{\textbf{k}}) - E_m(\textrm{\textbf{k}})} \frac{\left\langle n\left| \hat{v}_p\right| m\right\rangle\left\langle m\left| \hat{v}_q\right| n\right\rangle}{\hbar\omega + E_n(\textrm{\textbf{k}}) - E_m(\textrm{\textbf{k}}) + i\eta}&
\end{eqnarray}
where $S$ is the area of the primitive cell, $f(E)$ is the Fermi distribution function, $E_{n,m}(\textrm{\textbf{k}})$ and $\left| n,m\right\rangle$ represent the eigenenergies and eigenstates of the system, $\hat{v}_{p/q} = [\hat{r}_{p/q},\hat{H}] e/i\hbar$ are the velocity operators, and $\eta$ is a phenomenological broadening.

\begin{center}
	\includegraphics[width=0.82\columnwidth]{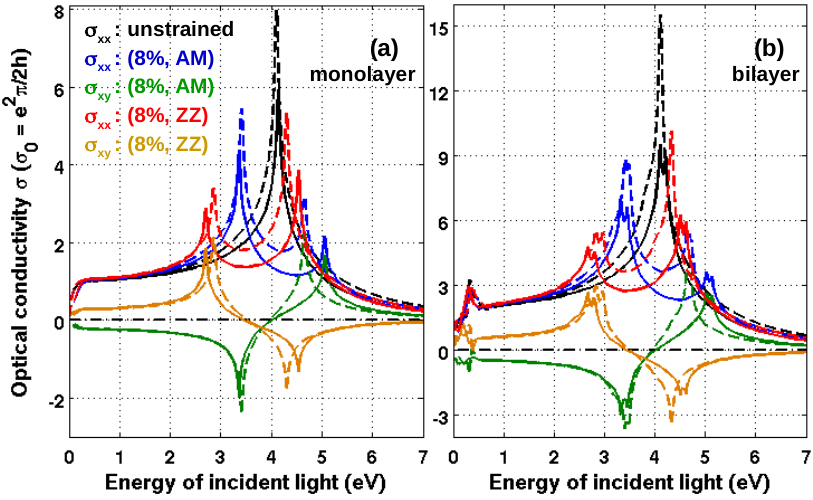} \\
	\textbf{Fig. S2}: Optical conductivity in (a) graphene monolayer and (b) bilayer systems under a uniaxial strain $\epsilon = 8\%$ applied in two different directions $\theta = 0^\circ$ (AM: armchair) and $90^\circ$ (ZZ: zigzag). The dashed and solid lines represent the DFT and TB results, respectively. The light polarization is fixed to $\phi = -45^\circ$.
\end{center}

In Figs.S2.a-b, the optical conductivity components are computed as a function of light energy in graphene monolayer and bilayer systems. Indeed, in all cases, our parametrized TB calculations reproduce very well the DFT results at low and high energies. A slight discrepancy between two methods occurs only in the medium energy range where the conductivity peaks are present. In spite of this fact, the two methods are still in very good agreement for the investigation of the overall spectrum of optical conductivities in both unstrained and strained graphene systems.\\

\begin{center}
	\includegraphics[width=0.82\columnwidth]{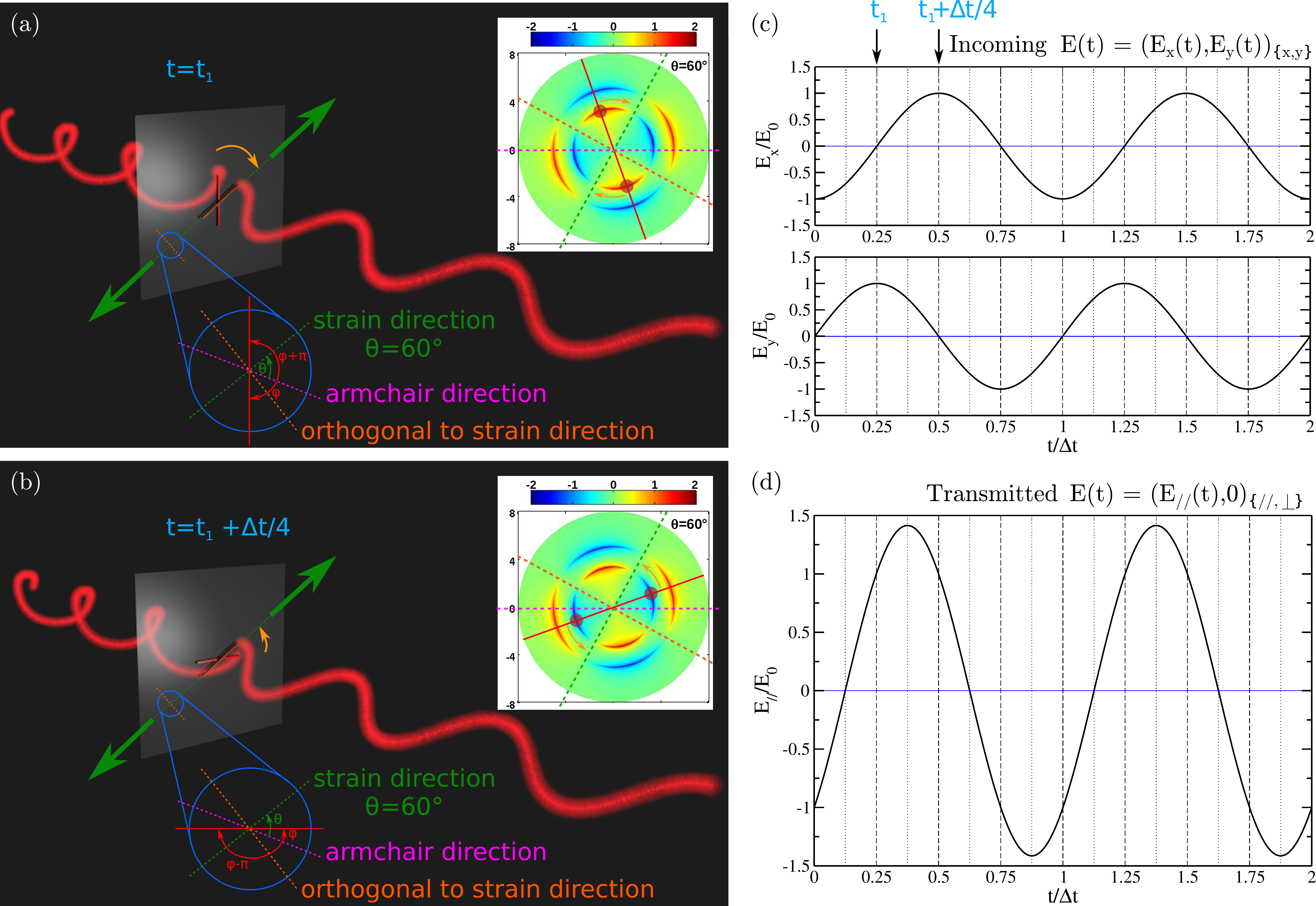} \\
    \textbf{Fig. S3}: Faraday rotation of a circularly polarized light reaching graphene layer at different times (a,b). The incoming circularly polarized light can be considered as a combination of two orthogonal linearly polarized lights (c) and in the low energy regime, the transmitted light is expected to be linearly polarized along the axis parallel to the strain (d).
\end{center}
\section{II. Faraday rotation of a circularly light}

In this section, we would like to present an illustration for the Faraday rotation of circularly polarized lights passing our considered systems. First, we would like to remind that the rotation angle $\delta \phi$ (resp. its direction) of a linearly polarized light depends on the value (resp. sign) of the Hall conductivity $\sigma_{xy}$, that has been demonstrated to be tunable by playing with different parameters such as incoming light energy ($\hbar \omega$), polarization angle ($\phi$), as well as strain strength ($\epsilon$) and type (uniaxial or shear strain), strain direction ($\theta$) and number of layers. Hence, as already discussed in the main text, an incoming linearly polarized light can be rotated differently, depending on the parameters mentioned above.

Now we consider the case of circularly polarized lights, which can be considered as a combination of two orthogonal linearly polarized lights (see Fig.S3.c) with a dephasing in time of a quarter the wave period ($\Delta t$). For a low energy light, both components rotates towards the strain axis either with a clockwise or anti-clockwise rotation as a function of time (see the illustration in Figs.S3.a-b). Hence, the transmitted light is expected to be linearly polarized along the strain axis (see Fig.S3.d) with the same period but an enhanced electric field. Thus, this device acts as a converter of circularly polarized light into a linearly polarized light.

Similarly, the transmitted light becomes linearly polarized along the orthogonal axis for a high energy incoming light (see also the related discussions in the main text). \\

\textbf{References}

$[1]$ J. M. Soler \textit{et al.}, J. Phys.Condens. Matter \textbf{14}, 2745 (2002).

$[2]$ J. P. Perdew \textit{et al.}, Phys. Rev. Lett. \textbf{77}, 3865 (1996).

$[3]$ N. Troullier and J. L. Martins, Phys. Rev. B \textbf{43}, 1993 (1991).

$[4]$ A. Lherbier \textit{et al.}, Phys. Rev. B \textbf{86}, 075402 (2012).

$[5]$ E. McCann and M. Koshino, Rep. Prog. Phys. \textbf{76}, 056503 (2013).

$[6]$ V. M. Pereira \textit{et al.}, Phys. Rev. B \textbf{80}, 045401 (2009).

$[7]$ O. L. Blakslee \textit{et al.}, J. Appl. Phys. \textbf{41}, 3373 (1970).

$[8]$ P. Moon and M. Koshino, Phys. Rev. B \textbf{87}, 205404 (2013).

$[9]$ H. A. Le \textit{et al.}, J. Phys.: Condens. Matter \textbf{26}, 405304 (2014).


\begin{thebibliography}{99}
\bibitem{hall79} E. H. Hall, Amer. J. Math. \textbf{2}, 287-292 (1879).
\bibitem{prgi90} R. E. Prange and S. M. Girvin, \textit{The Quantum Hall Effect, second edition} (Springer-Verlag, New York, 1990).
\bibitem{schu16} M. Schubert \textit{et al.}, J. Opt. Soc. Am. A \textbf{33}, 1553-1568 (2016).
\bibitem{karg15} M. Kargarian \textit{et al.}, Sci. Rep. \textbf{5}, 12683 (2015).
\bibitem{lebi11} L. Bi \textit{et al.}, Nat. Photon. \textbf{5}, 758-762 (2011).
\bibitem{jsch16} J. Schurr, IEEE Trans. Instrum. Meas. \textbf{19}, 24 - 26 (2016).

\bibitem{ferr15} A. C. Ferrari \textit{et al.}, Nanoscale \textbf{7}, 4598-4810 (2015).
\bibitem{jang16} H. Jang \textit{et al.}, Adv. Mater. \textbf{28}, 4184-4202 (2016). 
\bibitem{bona10} F. Bonaccorso \textit{et al.}, Nat. Photon. \textbf{4}, 611-622 (2010); Z. Sun \textit{et al.}, \textbf{ibid. 10}, 227-238 (2016).
\bibitem{chzi16} C. Si \textit{et al.},  Nanoscale \textbf{8}, 3207-3217 (2016).
\bibitem{kopp14} F. H. L. Koppens \textit{et al.}, Nat. Nanotechnol. \textbf{9}, 780-793 (2014).

\bibitem{grun03} A. Gr\"{u}neis \textit{et al.}, Phys. Rev. B \textbf{67}, 165402 (2003).
\bibitem{nair08} R. R. Nair \textit{et al.}, Science \textbf{320}, 1308 (2008). 
\bibitem{stau08} T. Stauber \textit{et al.}, Phys. Rev. B \textbf{78}, 085432 (2008).
\bibitem{yang09} L. Yang \textit{et al.}, Phys. Rev. Lett. \textbf{103}, 186802 (2009).
\bibitem{wrig09} A. R. Wright \textit{et al.}, Appl. Phys. lett. \textbf{95}, 163104 (2009); Phys. Rev. Lett. \textbf{103}, 207401 (2009); Nanotechnol. \textbf{20}, 405203 (2009).
\bibitem{yang10} C. H. Yang \textit{et al.}, Phys. Rev. B \textbf{82}, 205428 (2010).
\bibitem{fmak11} K. F. Mak \textit{et al.}, Phys. Rev. Lett. \textbf{106}, 046401 (2011).
\bibitem{yuan11} S. Yuan \textit{et al.}, Phys. Rev. B \textbf{84}, 195418 (2011).
\bibitem{moon13} P. Moon and M. Koshino, Phys. Rev. B \textbf{87}, 205404 (2013); \textbf{ibid.} \textbf{88}, 241412(R) (2013).
\bibitem{hale14} H. A. Le \textit{et al.}, J. Phys.: Condens. Matter \textbf{26}, 405304 (2014).
\bibitem{yang16} K. Yang \textit{et al.}, Sci. Rep. \textbf{6}, 23897 (2016).

\bibitem{mori09} T. Morimoto \textit{et al.}, Phys. Rev. Lett. \textbf{103}, 116803 (2009); Phys. Rev. B \textbf{86}, 155426 (2012).
\bibitem{cras11} I. Crassee \textit{et al.}, Nat. Phys. \textbf{7}, 48-51 (2011).
\bibitem{shim13} R. Shimano \textit{et al.}, Nat. Commun. \textbf{4}, 1841 (2013).
\bibitem{soun13} D. L. Sounas \textit{et al.}, Appl. Phys. Lett. \textbf{102}, 191901 (2013).
\bibitem{heym14} J. N. Heyman \textit{et al.}, J. Appl. Phys. \textbf{116}, 214302 (2014).
\bibitem{skul15} H. S. Skulason \textit{et al.}, Appl. Phys. Lett. \textbf{107}, 093106 (2015).
\bibitem{schi16} J. Schiefele \textit{et al.}, Phys. Rev. B \textbf{94}, 035401 (2016).
\bibitem{kuzm16} D. A. Kuzmin \textit{et al.}, Nano Lett. \textbf{16}, 4391-4395 (2016). 

\bibitem{pere10} V. M. Pereira \textit{et al.}, Euro. Phys. Lett. \textbf{92}, 67001 (2010).
\bibitem{pell10} F. M. D. Pellegrino \textit{et al.}, Phys. Rev. B \textbf{81}, 035411 (2010).
\bibitem{dong14} B. Dong \textit{et al.}, Nanotechnol. \textbf{25} 455707 (2014).
\bibitem{gxni14} G.-X. Ni \textit{et al.}, Adv. Mater. \textbf{26}, 1081-1086 (2014).
\bibitem{leyy15} M. Oliva-Leyva and G. G. Naumis, 2D Mater. \textbf{2}, 025001 (2015); Phys. Rev. B \textbf{93}, 035439 (2016).

\bibitem{supmat} See Supplemental Material for the detailed description of calculation methodologies and the illustration of Faraday rotation of a circularly polarized light.

\bibitem{blak70} O. L. Blakslee \textit{et al.}, J. Appl. Phys. \textbf{41}, 3373 (1970).
\bibitem{lher16} A. Lherbier \textit{et al.}, 2D Materials \textbf{3}, 045006 (2016).
\bibitem{sole02} J. M. Soler \textit{et al.}, J. Phys.: Condens. Matter \textbf{14}, 2745 (2002).
\bibitem{lher12} A. Lherbier \textit{et al.}, Phys. Rev. B \textbf{86}, 075402 (2012).
\bibitem{pere09} V. M. Pereira \textit{et al.}, Phys. Rev. B \textbf{80}, 045401 (2009).

\end{thebibliography}
\end{document}